# Hybrid Crossbar Architecture for a Memristor Based Memory

Chris Yakopcic, *Student Member*, *IEEE* and Tarek M. Taha, *Member*, *IEEE*

*Abstract*—This paper describes a new memristor crossbar architecture that is proposed for use in a high density cache design. This design has less than 10% of the write energy consumption than a simple memristor crossbar. Also, it has up to 4 times the bit density of an STT-MRAM system and up to 11 times the bit density of an SRAM architecture. The proposed architecture is analyzed using a detailed SPICE analysis that accounts for the resistance of the wires in the memristor structure. Additionally, the memristor model used in this work has been matched to specific device characterization data to provide accurate results in terms of energy, area, and timing.

## I. Introduction

AS CMOS devices have shrunk into the nanoscale regime, the increase in power density in CMOS systems has stopped the increase in single core processor performance. For this reason CPUs are now based on multicore architectures. There are two main factors that limit the performance of these architectures. First, there is currently not enough on-chip memory to effectively handle the instruction and data load that the multicore architecture is capable of processing. Second, power consumption limits the number of cores and on-chip memory, thus limiting performance.

As an alternative to traditional SRAM, Resistive Random Access Memory (RRAM) is a promising solution to the forthcoming memory wall problem in conventional CPUs. These memories work based on different resistive switching mechanisms where a dynamic resistance value determines the memory state of the device. The three main types of RRAM include memristors [1]-[2], Phase Change Random Access Memory (PCRAM) [3], and Spin-Torque Transfer Magnetic Random Access Memory (STT-MRAM) [4].

In 2008, the first physical realization of the memristor (initially theorized in 1971 [1]) was published [2]. Furthermore, memristor crossbar arrays have been proposed [5] as the potential building block of an ultra-high density memory system. The problem with these high density crossbar arrays is that the power consumption will increase dramatically with the size of the crossbar [6]. This is due to the many alternate current paths lowering the effective resistance of the array. Additionally, read errors are much more likely due to these alternate current paths. To solve this, a 1 transistor-1 memristor (1T1M) bit cell can be used which is commonplace in STT-MRAM architectures [7],[8]. Unfortunately, this will lower the density of the memristor memory system to that of a single transistor array.

This paper presents a memristor based memory system that is capable of achieving more than 4 times the density of a typical STT-MRAM array. Additionally, it has dramatically reduced power consumption when compared to a high-density transistor-less memristor crossbar. This is done by tiling many smaller memristor arrays for partially isolated resistive grids.

The analysis of this memory design is performed through SPICE simulation. To model the memristors, a previously published device model [9] is utilized that is capable of reproducing memristor characteristics very accurately. Both the wire resistance and the isolating transistors in the array are simulated to provide a more complete crossbar analysis. This work describes a very accurate device level simulation of a novel memristor based memory architecture, and provides results for energy consumption and noise margin within the circuit. An accurate area analysis is also performed that describes the layout of the memory system. Very few crossbar simulations [10],[11] account for wire resistance, and these were completed with less accurate device models.

This paper is organized as follows: Section II provides a comparison of the existing resistive memory devices and crossbar memory designs. Section III describes the design of the proposed memory architecture and Section IV discusses the procedure used to analyze the simulated crossbar tiles. Section V displays the results of the crossbar tile analysis and Section VI concludes the paper.

## II. Resistive Memory Technology

### A. Resistive Memory Devices

Resistive switching devices such as STT-MRAM, PCRAM, and memristor devices have all been proposed as possible solutions for the development of high density memory. These different types of resistive memory devices are used in a similar manner, although their properties differ slightly.

Manuscript received January 16, 2013. This work was supported by NSF CAREER Grant CCF-1053149.
C. Yakopcic is with the University of Dayton, Dayton, OH 45469, USA (email: cyakopcic1@udayton.edu).
T. M. Taha is with University of Dayton, Dayton, OH 45469 USA (phone: 937-229-3119; email: ttaha1@udayton.edu).
G. Subramanyam is with University of Dayton, Dayton, OH 45469 USA (email: gsubramanyam1@udayton.edu).

Previous research suggests [8][12] that STT-MRAM is the most promising candidate for the future of high-density, non-volatile, resistance switching memories. The $R_{OFF}/R_{ON}$ ratio of STT-MRAM is typically only about 2.5 [4],[13], so it is not likely that an STT-MRAM memory system would work without an access transistor for each individual memory device. This creates a problem where the maximum areal density of this type of memory system is limited by the size of an access transistor (similar to Fig. 1(a)) and not the nanoscale magnetic switching element.

PCRAM is another promising new memory technology. It has the lowest endurance in terms of switching cycles before failure, and generally has a longer switching time (50 to 100ns) [14] when compared to memristors and STT-MRAM. For these reasons, the majority of the PCRAM based memory systems are proposed as a replacement for DRAM as opposed to SRAM. However, PCRAM has the advantage of unipolar switching [3], so diodes can be used to limit unwanted current paths in a higher density design.

The memristor device that was selected for the final results in this paper has a relatively fast switching time (10ns) and very low current draw with an on state resistance of 125kΩ [5]. Additionally the device has a very large off to on ratio (about $10^6$) that will be very useful in the proposed design since a limited number of unwanted current paths will be present. A number of other memristor devices [15],[16] were tested for use in the system, but according to our simulations they either had a power consumption that was too large, or a $R_{OFF}/R_{ON}$ ratio that was too small.

### B. Crossbar Array Designs

A common solution to eliminate alternate current paths in a resistive memory system is to place an access transistor alongside each memory element (see Fig. 1(a)). This technique greatly reduces the chance of a read error and limits the power consumptions since greater control is placed on the path of the current flow. The disadvantage of this type memory system is that the areal density of the system is now limited by the area of the transistors and not the area of the memory devices.

The circuit diagram and layout for a high density memory crossbar can be seen in Fig. 1(b) and (c). In this design nanoscale memory elements can be packed at a much higher density. Each memory element will consume an area of just $4F^2$ [17] where $F$ is the minimum feature size of the fabrication technique. The schematic in Fig. 1(b) illustrates the problem with this type of crossbar. When voltages are applied to the wires $a_1$ and $b_1$, nothing is stopping current from flowing through other devices. This can lead to read errors in a large crossbar because the current sensed at $b_1$ could be due to a chain of devices in a low resistance state when the selected memristor is in a high resistance state. This also increases the power consumption of the crossbar, as the current draw increases due to the many alternate current paths.

Some preliminary simulation results show how the energy consumption of a high density crossbar increases with crossbar size in Fig. 1(d). The value plotted is the energy required to write a single bit. Since the simulations were performed in SPICE, a 16×16 crossbar containing 256 memristors was the largest system that could be simulated. If assumed to be linear, the data can be extrapolated to show that large crossbars quickly reach a level of energy consumption that that would be unrealistic for a competitive memory technology. This study was performed assuming a 500Ω wire resistance between all memristors modeled after the device in [5] using 10ns ±7V write and erase pulses.

As a possible solution to this problem, one publication [18] presents a memristor device with current suppression in one direction. This reduces the problem of alternate current paths, although the switching time of this device is too large for an on-chip memory application (100μs).

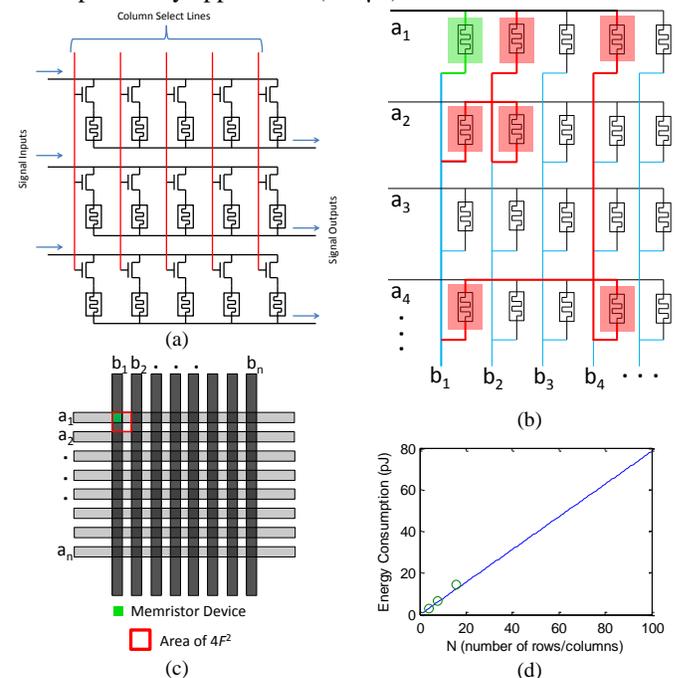

Fig. 1. Memristor crossbars including (a) a 1T1M architecture schematic, and (b) a high density unconstrained crossbar displaying the target memristor path (green) as well as two possible alternate current paths (red). For the circuit in (b) we show the (c) layout, and (c) write energy consumption.

### III. PROPOSED HYBRID CROSSBAR DESIGN

The proposed hybrid crossbar architecture is a combination between a high density array, and one with transistor isolation. In this design, transistors are used to isolate small crossbars within a larger array. These smaller memristor arrays will be referred to as tiles. Fig. 2 displays a portion of the circuit design for the hybrid memory system.

In this example, 4 memristor tiles are displayed, each consisting of 16 memristors arranged into a 4×4 square. The top of the circuit displays a pulse generator block, which is responsible for sending data to a single row in each tile (through either $D_{R1}$, $D_{R2}$, $D_{R3}$, or $D_{R4}$) and grounding the rest. Additionally, a row decoder containing the row select signals ($S_1$ through $S_N$) is designed to turn on only one row of tiles during a parallel read or write operation. During a read or write operation, data from the selected row of tiles will be processed by the column circuits. This design allows for all unwanted current paths to be contained within the small 4×4 crossbars with twice the bit-cell density of a 1T1M design.

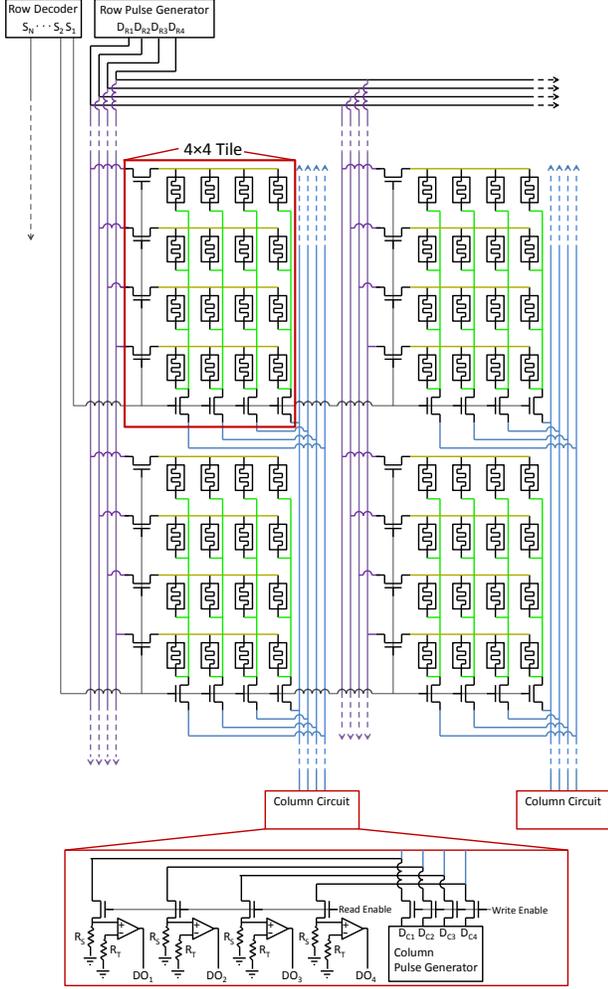

Fig. 2. Circuit diagram for the tiled crossbar memory system.

A write operation in this design is a two-step process [19] that writes to an entire row of memristors in each of the selected tiles (see Fig. 3). The first step is to apply a voltage of $V_w/2$ to a selected row (grounding the other 3), and to apply a voltage of $-V_w/2$ to all columns where a 1 (low resistance state) should be written (where $V_w$ is the write voltage). Furthermore, a voltage of $V_w/2$ should be applied to all global column wires were a 0 (high resistance state) should be written (through the write enable transistors). This will result in writing a 1 to only the memristors in the selected row that need to be set to 1. During the second step in the write process, a voltage of $-V_w/2$ is applied to the selected row, and all global column wires are set as they were in step one. This will result in writing a 0 to the rest of the memristors in the row.

A parallel read operation is performed by setting a selected row of memristors to a voltage below the switching threshold, and activating the read enable transistors in each column circuit. The analog read voltage across $R_S$ (in Fig. 2) is converted to a binary signal using a comparator and the constant threshold resistance $R_T$.

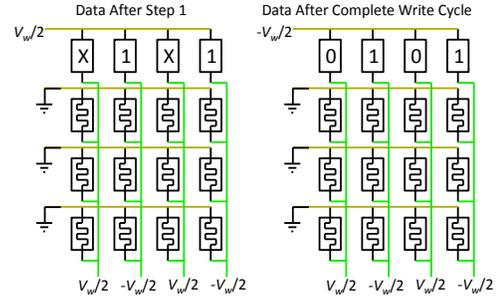

Fig. 3. Demonstration of the write operation on a crossbar tile.

## IV. HYBRID CROSSBAR ANALYSIS

### A. Memristor Device Model

To perform a device level analysis of a memristor crossbar memory system, a SPICE equivalent of the memristor model first proposed in [9] was utilized. This model was set to match the characterization data of one of the memristor devices published in [5] (see Fig. 4). This device was chosen for use in the proposed memory design because it had a large $R_{OFF}/R_{ON}$ ratio ($10^6$) while still retaining a relatively low switching time (about 10 ns). It also has a large on state resistance of about 125kΩ (determined by the 8µA current from a 1V read pulse).

The simulation result in Fig. 4 shows the minimum and maximum resistances of the model to be 124.95kΩ and 125.79×10$^9$Ω respectively, which correlates very closely to the characterization [5]. Applying a +7V pulse successfully switches the device into a low resistance state, and applying a -7V pulse drives the model into a high resistance state. These strong simulation results show that a reliable device model has been developed, and this will lead to more accurate results when simulating memristors.

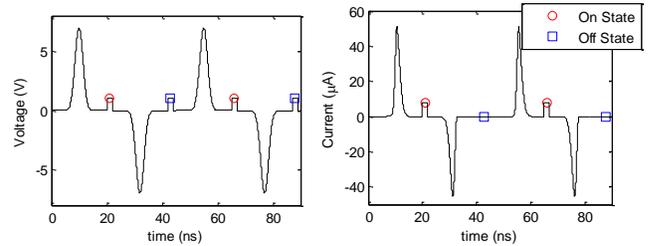

Fig. 4. Simulation results displaying the input voltage and current waveforms. The following parameter values were used in the model [5] to obtain this result: $V_p$=1.088V, $V_n$=1.088V, $A_p$=816000, $A_n$=816000, $x_p$=0.985, $x_n$=0.985, $α_p$=0.1, $α_n$=0.1, $a_1$=1.6(10-4), $a_2$=1.6(10-4), $b$=0.05, $x_0$=0.01.

### B. SPICE Circuit Simulation

To determine the maximum noise margin and energy consumption of both the 4×4 and 8×8 tiles, a large string of read and write signals was applied to a crossbar simulation. Crossbar circuit operation varies based on the resistance values of memristors within the crossbar, so a large number of randomized signals were applied to obtain an average result.

To complete this task, the signals were generated in MATLAB and then saved in a text file that could be interpreted in LTSpice (see Fig. 5). The signals generated included switching signals for the transistors to select the correct memristors, as well as the data signals that contained the read and write pulses.

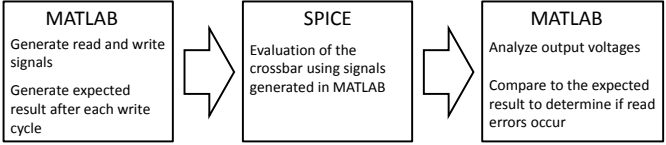

Fig. 5. Block diagram for MATLAB/SPICE simulation process.

The row of memristors that was to be written was chosen at random by the MATLAB script. The data that was written in the memristors was also randomized by choosing to apply either a +7 or -7V pulse signifying a write or erase operation respectively. These voltages were chosen to match the switching characteristics in [5]. After each write operation was performed, a read operation was performed that read each of the memristors in the crossbar one row at a time.

A 3-dimensional binary answer matrix (*x*-dimension: crossbar row, *y*-dimension: crossbar column, *z*-dimension: simulation cycle) was also generated at the time of the random signal creation in MATLAB. This answer matrix held the values expected to be obtained from a read after each write operation was performed. This was compared with the analog read data (the voltage across sense resistors) to determine the maximum error free noise margin.

To perform this task, the analog read voltages ($V_s$) across the sense resistors ($R_s$) were imported into MATLAB and the voltage peaks at each read were extracted. These values were compared to the data in the answer matrix as the dead zone between 1 and 0 was increased. The maximum width of the dead zone (in mV) that did not produce a read error was considered to be the noise margin (see Fig. 6). It should be noted that previous publications [19] have proposed methods for determining the sense resistance value to produce maximum noise margin. However these methods did not hold true in our case most likely due to the added wire resistance.

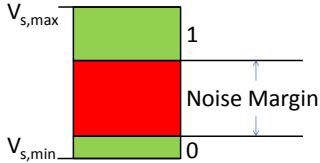

Fig. 6. Noise margin in memristor crossbars.

To determine the write energy, a large number of write pulses (without reading afterward) were applied to the crossbar tile. The total energy consumption in the circuit was then divided by the number of writes to determine the average write energy per bit. Determining the read energy was a similar process where a series of reads (without writes) was applied to the crossbar initialized with a random data pattern to determine the average read energy per bit.

V. CROSSBAR TILE RESULTS

*A. Energy Analysis*

A large number of simulations were completed to determine the optimal sense resistance and write voltage that would maximize the noise margin (see Table 1) for the 4×4 and 8×8 crossbar tiles. When considering transistor and high nano-wire resistances [20], write errors can be a more common occurrence even when using a write technique [19] thought to eliminate write errors. Write voltage must be increased in the presence of wire resistance to ensure the selected memristor devices would be fully switched. However, increasing the write voltage also increases the probability that a voltage drop across a half-selected device (see Fig. 3) will be greater than the memristor write threshold, leading to unwanted changes in the stored data. When comparing the two tile designs, Table 1 shows that the 8×8 crossbar provides twice the bit density. Although, it has a lower noise margin and consumes more energy due to the alternate current paths.

Table 1. Comparison of the 4×4 and 8×8 tiles with a 500Ω wire resistance between all devices.

| Tile Size (Ω) | 4×4 | 8×8 |
|---|---|---|
| Write/Erase Voltage (V) | 7 | 7.5 |
| Write Energy (pJ) | 3.1176 | 6.3722 |
| Read Energy (fJ) | 5.0561 | 6.8491 |
| Opt. Sense Resistance (Ω) | 8Ron | 0.5Ron |
| Max. Noise Margin (mV) | 410 | 66 |
| Wire Resistance (Ω) | 500 | 500 |

Table 2 compares the proposed memory system to other memory architectures. The values for bit density were calculated assuming 45nm technology for the memristor and SRAM options. The feature size of the STT-MRAM bit cells used was 65nm [8]. The 4×4 and 8×8 tiled systems consume less than 5% and 10% of the energy consumed in the unconstrained 1kB crossbar respectively. Furthermore, the read energy in the tiled systems is less than 1/3 of the read energy in the unconstrained crossbar.

The proposed architecture has a higher density when compared to SRAM, although it has a longer write time. Also, SRAM has a large amount of leakage energy [21] that is not present in the memristor based systems. This is because memristors do not require power to retain their memory state, and the transistors in the crossbar tiles are only active when a read or write pulse is present.

The SRAM leakage current was assumed to be 29μA/Mb with a 1V operating voltage [21]. To obtain leakage energy of a single-bit access, it was assumed that an average of 1000 accesses per bit would be performed in one second. It should be noted that the leakage will be consumed by all memory cells within an SRAM array even if they are not being accessed, so this will lead to larger total energy consumption.

Table 2. Performance comparison of different memory cell designs.

| Memory Architecture | SRAM Active | SRAM Leakage | STT-MRAM | Hybrid (4×4) | Hybrid (8×8) | 1kB Crossbar |
|---|---|---|---|---|---|---|
| Bit Density (Gbits/cm²) | 0.338 | | 0.760 | 1.98 | 3.95 | 12.35 |
| Read Energy (fJ/bit) | 0.7 | 27.7 | 60.4 | 5.506 | 6.849 | 21 |
| Write Energy (fJ/bit) | 0.7 | 27.7 | 1177 | 3118 | 6372 | 70000 |
| Read Time (ns) | 0.3 | | 0.3 | 0.3 | 0.3 | 0.3 |
| Write Time (ns) | 0.3 | | 0.57 | 10 | 10 | 10 |

*B. Area Analysis*

If the system utilizes 4×4 tiles, then it will require 8 transistors to drive a tile consisting of 16 memristors (an average of 2 memristors per transistor). This system was also analyzed where each of the tiles consisted of 64 memristors in an 8×8 arrangement. The 8×8 tile consists of 16 transistors and 64 memristors (an average of 4 memristors per transistor). In these designs the size of the transistors is still the limiting factor in memory density. However, the 4×4 and 8×8 tile systems provide an increase in density over a 1T1M system by a factor of 2 and 4 respectively. The layouts for the 4×4 and 8×8 tiles can be seen in Fig. 7, where the red squares represent memristors. Standard CMOS design rules limit the transistor cell size to $50F^2$. For the memristor layers, nanoscale fabrication methods should be able to pack memristors at a higher density [2],[18] than standard fabrication methods.

The peak drain current of any single transistor during a write in the 4×4 tile was about 120μA, and about 200μA in the 8×8 tile. These currents are significantly lower than the requirement for energy efficient STT-MRAM cells [4],[8]. For this reason the transistor packing density in the hybrid array can be the same or higher than that of STT-MRAM.

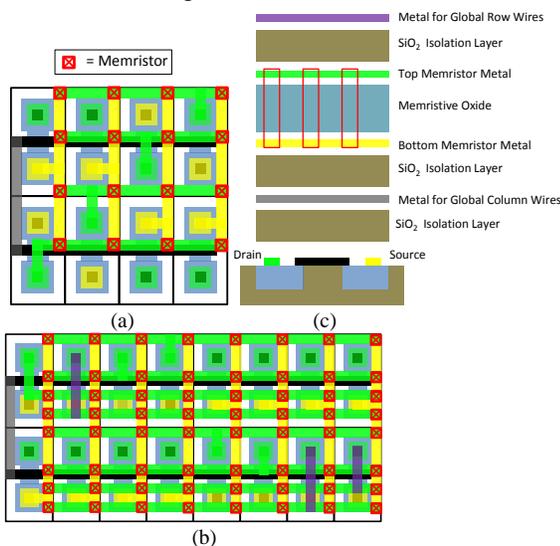

Fig. 7. Layout for (a) the 4×4 tile, (b) the 8×8 tile, and (c) the layer organization for each of the tiles.

## VI. Conclusion

A new hybrid memory system has been proposed that can provide up to 5.2 times the memory density of an STT-MRAM system. The proposed system has a significantly lower energy consumption compared to a large high density memristor crossbar. A detailed analysis of the design including wire resistance and accurate device modeling was performed in SPICE. Future work includes a further study of the memory system to see if larger tiles would benefit the system. Our existing simulations show that a 16×16 tile is not capable of producing a significant noise margin, although it may be possible to correct this by modeling alternative memristor devices. If larger tiles are used, it may be possible for this system to approach the bit density of a transistor-less crossbar.

## References

[1] L. O. Chua, "Memristor—The Missing Circuit Element," *IEEE Transactions on Circuit Theory*, 18(5), 507–519 (1971).
[2] D. B. Strukov, G. S. Snider, D. R. Stewart, and R. S. Williams, "The missing Memristor found," *Nature*, 453, 80–83 (2008).
[3] H.-S. P. Wong, S. Raoux, S. Kim, J. Liang, J. P. Reifenberg, B. Rajendran, M. Asheghi, and K. E. Goodson, "Phase Change Memory," Proceedings of the IEEE, vol. 98, no. 12, pp. 2201-2207, Dec. 2010.
[4] Z. M. Zeng, P. Khalili Amiri, G. Rowlands, H. Zhao, I. N. Krivorotov, J.-P. Wang, J. A. Katine, J. Langer, K. Galatsis, K. L. Wang, and H. W. Jiang, "Effect of resistance-area product on spin-transfer switching in MgO-based magnetic tunnel junction memory cells," Appl. Phys. Lett. 98, 072512 (2011).
[5] W. Lu, K.-H. Kim, T. Chang, S. Gaba, "Two-terminal resistive switches (memristors) for memory and logic applications," in Proc. 16$^{th}$ Asia and South Pacific Design Automation Conference, 2011, pp. 217-223.
[6] S. Shin, K. Kim, S.-M. Kang, "Analysis of Passive Memristive Devices Array: Data-Dependent Statistical Model and Self-Adaptable Sense Resistance for RRAMs" Proceedings of the IEEE, 100(6), June 2012.
[7] C. J. Lin et al., "45nm low power CMOS logic compatible embedded STT MRAM utilizing a reverse-connection 1T/1MTJ cell," *IEDM*, 2009
[8] X. Guo, E. Ipek, and T. Soyata. "Resistive Computation: Avoiding the Power Wall with Low-Leakage LUT-Based Computing," ISCA, (2010).
[9] C. Yakopcic, T. M. Taha, G. Subramanyam, R. E. Pino, and S. Rogers, "A memristor device model," *IEEE Electron Device Lett*., vol. 32, no. 10, pp. 1436-1438, Oct. 2011.
[10] O. Kavehei, S. Al-Sarawi, K.-R. Cho, K. Eshraghian, and D. Abbott, "An Analytical Approach for Memristive Nanoarchitectures," IEEE Trans. on Nanotechnology 11(2), March 2012, 374-385.
[11] A. Heittmann and T. G. Noll, "Limits of Writing Multivalued Resistances in Passive Nanoelectronic Crossbars Used in Neuromorphic Circuits," GLSVLSI 2012 227-232.
[12] A. K. Mishra, X. Dong, G. Sun, Y. Xie, N. Vijaykrishnan, and C. R. Das, "Architecting on-chip interconnects for stacked 3D STT-RAM caches in CMPs," *Proceeding of the 38th annual international symposium on Computer architecture*, June 04-08, 2011, San Jose, CA.
[13] G. Guo, X. Guo, Y. Bai, and E. Ipek. "A Resistive TCAM Accelerator for Data Intensive Computing," *International Symposium on Microarchitecture (MICRO)*, Porto Alegre, Brazil, November 2011.
[14] M. Zhu, L. Wu, Z. Song1 F. Rao, D. Cai, C. Peng, X. Zhou, K. Ren, S. Song, B. liu, and S. Feng, "$Ti_{10}Sb_{60}Te_{30}$ for phase change memory with high-temperature data retention and rapid crystallization speed," Applied Physics Letters, 100(12), 2012.
[15] J. P. Strachan, A. C. Torrezan, G. Medeiros-Ribeiro, and R. S. Williams, "Measuring the switching dynamics and energy efficiency of tantalum oxide memristors," Nanotechnology, 22(50), (505402), Dec. 2011.
[16] F. Miao, J. P. Strachan, J. J. Yang, M.-X. Zhang, I. Goldfarb, A. C. Torrezan, P. Eschbach, R. D. Kelley, G. Medeiros-Ribeiro, R. S. Williams, 'Anatomy of a Nanoscale Conduction Channel Reveals the Mechanism of a High-Performance Memristor," Advanced Materials, vol. 23, no. 47, pp. 5633-5640, Nov. 2011.
[17] K. Eshraghian, K. –R. Cho, O. Kavehei, S. –K. Kang, D. Abbott, and S. –M. S. Kang, "Memristor MOS Content Addressable Memory (MCAM) Hybrid Architecture for Future High Performance Search Engines," IEEE Trans. On VLSI Systems, vol. 19, no. 8, 1407–1417, Aug. 2011.
[18] K.-H. Kim, S. Gaba, D. Wheeler, J. M. Cruz-Albrecht, T. Hussain, N. Srinivasa, and W. Lu, "A Functional Hybrid Memristor Crossbar-Array/CMOS System for Data Storage and Neuromorphic Applications," Nano Letters, 12(1), 2012, pp. 389-395.
[19] X. Dong, C. Xu, S. Member, Y. Xie, and N. P. Jouppi, "NVSim: A Circuit-Level Performance, Energy, and Area Model for Emerging Nonvolatile Memory," IEEE Trans. on Computer Aided Design of Integrated Circuits and Systems, 31(7), July, 2012, pp. 994-1007.
[20] N. S. Malvankar, M. Vargas, K. P. Nevin, A. E. Franks, C. Leang, B.-C. Kim, K. Inoue, T. Mester, S. F. Covalla, J. P. Johnson, V. M. Rotello, M. T. Tuominen and D. R. Lovley, "Tunable metallic-like conductivity in microbial nanowire networks," Nature Nanotech., 6, (2011), 573-579.
[21] N. Maeda, S. Komatsu, M. Morimoto, K. Tanaka, Y. Tsukamoto, K. Nii, and Y. Shimazaki, "A 0.41 μA Standby Leakage 32 kb Embedded SRAM with Low-Voltage Resume-Standby Utilizing All Digital Current Comparator in 28 nm HKMG CMOS," IEEE Jounal of Solid State Physics, 48(4), April 2013.